\documentclass[secnumarabic,amssymb,nobibnotes,notitlepage,superscriptaddress,11pt]{revtex4-1}
\usepackage{hyperref}
\usepackage{graphics}
\usepackage{graphicx}
\usepackage{epspdfconversion}

\newcommand{\ket}[1]{| #1 \rangle}

\begin{document}
\title{Controlling the coherence of a diamond spin qubit through strain engineering}%
\author{Young-Ik Sohn}%
\thanks{These authors contributed equally}
\author{Srujan Meesala}
\thanks{These authors contributed equally}
\affiliation{John A. Paulson School of Engineering and Applied Sciences, Harvard University, 29 Oxford Street, Cambridge, MA 02138, USA}
\author{Benjamin Pingault}
\thanks{These authors contributed equally}
\affiliation{Cavendish Laboratory, University of Cambridge, J. J. Thomson Avenue, Cambridge CB3 0HE, UK}
\author{Haig A. Atikian}
\affiliation{John A. Paulson School of Engineering and Applied Sciences, Harvard University, 29 Oxford Street, Cambridge, MA 02138, USA}
\author{Jeffrey Holzgrafe}
\affiliation{John A. Paulson School of Engineering and Applied Sciences, Harvard University, 29 Oxford Street, Cambridge, MA 02138, USA}
\affiliation{Cavendish Laboratory, University of Cambridge, J. J. Thomson Avenue, Cambridge CB3 0HE, UK}
\author{Mustafa G\"undo\u{g}an}
\affiliation{Cavendish Laboratory, University of Cambridge, J. J. Thomson Avenue, Cambridge CB3 0HE, UK}
\author{Camille Stavrakas}
\affiliation{Cavendish Laboratory, University of Cambridge, J. J. Thomson Avenue, Cambridge CB3 0HE, UK}
\author{Megan J. Stanley}
\affiliation{Cavendish Laboratory, University of Cambridge, J. J. Thomson Avenue, Cambridge CB3 0HE, UK}
\author{Alp Sipahigil}
\affiliation{Department of Physics, Harvard University, 17 Oxford Street, Cambridge, MA 02138, USA}
\author{Joonhee Choi}
\affiliation{John A. Paulson School of Engineering and Applied Sciences, Harvard University, 29 Oxford Street, Cambridge, MA 02138, USA}
\affiliation{Department of Physics, Harvard University, 17 Oxford Street, Cambridge, MA 02138, USA}
\author{Mian Zhang}
\affiliation{John A. Paulson School of Engineering and Applied Sciences, Harvard University, 29 Oxford Street, Cambridge, MA 02138, USA}
\author{Jose L. Pacheco}
\affiliation{Sandia National Laboratories, Albuquerque, NM 87185, USA}
\author{John Abraham}
\affiliation{Sandia National Laboratories, Albuquerque, NM 87185, USA}
\author{Edward Bielejec}
\affiliation{Sandia National Laboratories, Albuquerque, NM 87185, USA}
\author{Mikhail D. Lukin}
\affiliation{Department of Physics, Harvard University, 17 Oxford Street, Cambridge, MA 02138, USA}
\author{Mete Atat\"ure}
\affiliation{Cavendish Laboratory, University of Cambridge, J. J. Thomson Avenue, Cambridge CB3 0HE, UK}
\author{Marko Lon\v{c}ar}
\affiliation{John A. Paulson School of Engineering and Applied Sciences, Harvard University, 29 Oxford Street, Cambridge, MA 02138, USA}

\maketitle

\textbf{The uncontrolled interaction of a quantum system with its environment is detrimental for quantum coherence. In the context of solid-state qubits, techniques to mitigate the impact of fluctuating electric \cite{YiwenNanoLett, Shanying, IBM.glycerol} and magnetic fields \cite{balasubramanian, phosphorus.in.si, optimal.dd, dobrovitskii, DD.review} from the environment are well-developed. In contrast, suppression of decoherence from thermal lattice vibrations is typically achieved only by lowering the temperature of operation. Here, we use a nano-electro-mechanical system (NEMS) to mitigate the effect of thermal phonons on a solid-state quantum emitter without changing the system temperature. We study the silicon-vacancy (SiV) colour centre in diamond which has optical and spin transitions that are highly sensitive to phonons \cite{PhysRevLett.113.263601, PhysRevLett.113.263602, BeckerUltraFast, SiVMicrowave, SiVNJP}. First, we show that its electronic orbitals are highly susceptible to local strain, leading to its high sensitivity to phonons. By controlling the strain environment, we manipulate the electronic levels of the emitter to probe, control, and eventually, suppress its interaction with the thermal phonon bath. Strain control allows for both an impressive range of optical tunability and significantly improved spin coherence. Finally, our findings indicate that it may be possible to achieve strong coupling between the SiV spin and single phonons, which can lead to the realisation of phonon-mediated quantum gates \cite{SAWquantum} and nonlinear quantum phononics \cite{TahanPhonodynamics,Lodahl,Cleland,Chu.Transmon}.}

Phonons couple to solid-state emitters directly through periodic deformation of the electronic wavefunctions \cite{stoneham}. Electron-phonon interactions are responsible for relaxation and decoherence processes in a variety of quantum systems \cite{HansonReview, qd.phonon.decay, SiVNJP, fuchs.motional.narrowing, siyushev.gev, bhaskar.gev, orbach}. In particular, for systems with spin-orbit coupling, phonon-mediated processes can demand operation at sub-Kelvin temperatures \cite{SiV.fridge, faraon.cavity}, or the use of magnetic fields of several Tesla \cite{kramers.highBfield} to achieve long spin relaxation and coherence times. This requires cryogenic setups that are significantly more complex than common helium-4 cryostats employed to obtain coherent optical photons from solid state emitters. In contrast, our approach takes advantage of the fact that the large electron-phonon coupling responsible for such deocherence processes fundamentally arises from a high susceptibility of the electronic orbitals to lattice strain. We use this property to quench the effect of the thermal phonon bath on a single electronic spin qubit without lowering the operating temperature. Our experiments are performed on the negatively charged silicon-vacancy (SiV) centre in diamond, an emerging platform for photonic quantum networks \cite{SiVcqed} with remarkable optical properties owing to its inversion symmetry \cite{AlpHOM}. This inversion symmetry is also responsible for the particular electronic structure of the SiV, shown in Fig. 1a, with similar ground-state (GS) and excited-state (ES) manifolds, each containing two distinct orbital branches \cite{Hepp2014}. Orbital degeneracy in each manifold is lifted by spin-orbit coupling: $\ket{1},\ket{2}$ in the GS split by 46 GHz, and $\ket{3},\ket{4}$ in the ES split by 255 GHz in the absence of strain. Phonons with frequencies corresponding to these splittings can drive orbital transitions within the ground and excited manifolds \cite{SiVNJP}. 

As a first step towards controlling the electron-phonon interaction, we investigate the effect of static strain on these orbitals through  strain-dependent photoluminescence excitation (PLE) of the optical transitions labelled A, B, C and D at 4\,K. Static strain control at the location of the emitter is achieved with a NEMS device, a monolithic single-crystal diamond cantilever with metal electrodes patterned above and below it \cite{SuppInfo}, as shown in the scanning electron microscope (SEM) image in Fig. 1b. An opening in the top electrode allows optical access to SiV centres located in an array (inset of Fig. 1b), precisely positioned by focused ion-beam (FIB) implantation of $^{28}$Si$^{+}$ ions \cite{japan.fib, dirk.fib}. A DC voltage applied across the electrodes deflects the cantilever downwards due to electrostatic attraction and generates controllable static strain oriented predominantly along the long axis of the cantilever. The strain profile can be simulated numerically via the finite-element-method (FEM), as shown in Fig. 1c. Of the two possible orientations of SiVs in our device, we address those with transverse orientation (labelled blue, and shown in detail in inset of Fig. 1c), which predominantly experience strain in the plane normal to their highest symmetry axis ($E_{g}$ $-$ symmetric strain \cite{Hepp.thesis}). Upon applying strain, transitions A and D shift towards shorter and longer wavelengths, respectively. These shifts indicate increasing GS and ES splittings as shown in Fig. 2a. This result is consistent with a previous experiment on a dense ensemble of SiVs \cite{Sternschulte}. The variations in GS and ES splittings shown in Fig. 2a are quadratic at low strain, and linear at high strain. This indicates that $E_{g}$ $-$ symmetric strain mixes orbitals within the GS and ES manifolds, and thus phonon modes with corresponding strain components can induce resonant transitions between these orbitals. In contrast, strain along the SiV axis ($A_{1g}$ $-$ symmetric strain) is found to leave the GS and ES splittings unchanged, and therefore cannot cause electronic transitions. Complete characterisation of the strain response and relevant group theory analysis are detailed elsewhere \cite{SiVPRB}. 

With our device we can tune the splitting of the orbitals in the GS manifold from 46 GHz to typically up to 500 GHz, and in the best case, up to 1.2 THz \cite{SuppInfo}. In doing so, we can probe the interaction between SiV and the phonon bath at different frequencies by measuring the thermal relaxation rate of the orbital with a time-resolved pump-probe technique (Fig. 2b). Measurements are performed in the frequency range $\Delta_{\mathrm{gs}} =$ 46 GHz to 110 GHz where this technique can be applied. The total relaxation rate is a sum of the rates of phonon absorption, $\gamma_{\mathrm{up}}$, and emission, $\gamma_{\mathrm{down}}$ (shown in Fig. 1a), which can be individually extracted using the theory described in \cite{SuppInfo}. Over the range of $\Delta_{\mathrm{gs}}$ measured, phonon processes in both directions are observed to accelerate with increasing orbital splitting, thus indicating that the number of acoustic modes resonant with the GS splitting, i.e. the phonon density of states (DOS) at this frequency, increases with an expected dependence in $\Delta_{\mathrm{gs}}^n$ ($n$ depends on the geometry of material seen by resonant phonons \cite{SuppInfo}). However, if the orbital splitting is increased far above 120 GHz (at temperature $T$ = 4\,K) as plotted in Fig. 2c, the phonon absorption rate ($\gamma_{\mathrm{up}}$) is theoretically expected to reverse its initial trend. In this regime, the polynomial increase in phonon DOS is outweighed by the exponentially decrease in thermal phonon occupation ($\sim \mathrm{exp}(-h\Delta_{\mathrm{gs}}/k_BT)$) \cite{SiVNJP}, and consequently $\gamma_{\mathrm{up}}$ is rapidly quenched.

Such a suppression of phonon absorption at high strain can improve the spin coherence of the emitter. In the presence of magnetic field, the SiV electronic levels further split into spin sub-levels and provide an optically accessible spin qubit as shown in Fig. 3a \cite{PhysRevLett.113.263601, PhysRevLett.113.263602, Muller2014}. We use coherent population trapping (CPT) through simultaneous resonant laser excitation of the optical transitions labeled C1 and C2 to pump the SiV into a dark state, a coherent superposition of the spin sub-levels $\ket{1\downarrow}$, $\ket{1\uparrow}$. When the two-photon detuning is scanned, preparation of the dark state results in a fluorescence dip, whose linewidth is determined by the optical driving and spin dephasing rates. At low laser powers, the linewidth is limited by spin dephasing, which is dominated by phonon-mediated transitions within the GS manifold \cite{SiVNJP, SuppInfo}. In Fig. 3b, as the dark resonance narrows down due to prolonged spin coherence with increasing strain, we reveal a fine structure not visible before. Further measurements in Ref. \cite{SuppInfo} suggest that the presence of two resonances is due to interaction of the SiV electron spin with a neighbouring spin such as a $^{13}$C nuclear spin. This indicates the possibility of achieving a local register of qubits as has been demonstrated with nitrogen vacancy (NV) centres \cite{Childress13102006}. Fig. 3c shows the decreasing linewidths of the CPT resonances with increasing GS orbital splitting, indicating an improved spin coherence time. Beyond a GS splitting of $\sim$400 GHz, the linewidths saturate at $\sim$1 MHz. At the highest strain condition, we perform a power dependent CPT measurement to eliminate the contribution of power broadening, and extract a spin coherence time of $T_2^* = 0.25 \pm 0.02 \,\mathrm{\mu s}$ (compared with $T_2^*$ = 40\,ns without strain control \cite{PhysRevLett.113.263601, PhysRevLett.113.263602}). This saturation of $T_2^*$ suggests the mitigation of the primary dephasing source, single-phonon transitions between the GS orbitals, and the emergence of a secondary dephasing mechanism such as slowly varying magnetic fields from naturally abundant (1.1\%) $^{13}$C nuclear spins in diamond. Our longest $T_2^* = 0.25 \pm 0.02 \,\mathrm{\mu s}$ is on par with that of the NV center without dynamical decoupling \cite{balasubramanian, emre.cpt}, and of low-strain SiVs operated at a much lower temperature of 100 mK \cite{SiV.fridge}, the conventional approach to suppress phonon-mediated dephasing. 

In conclusion, we use a nano-electro-mechanical system to probe and control the interaction between a single electronic spin and the phonon bath of its solid-state environment. In doing so, we demonstrate six-fold prolongation of spin coherence by suppressing phonon-mediated dephasing as the dominant decoherence mechanism. As a next step, we can further improve the spin coherence by cancelling the effect of slowly-varying non-Markovian noise from the environment \cite{SiV.fridge} using dynamical decoupling techniques that are well-studied with other spin systems \cite{optimal.dd, dobrovitskii, Childress13102006}. Our strain engineering approach can be applied to overcome phonon-induced decoherence in other emitters such as emerging inversion-symmetric centers in diamond \cite{siyushev.gev, bhaskar.gev, iwasaki.snv, tchernij.snv}, Kramers rare earth ions \cite{orbach, faraon.cavity, kramers.highBfield}, and in general, systems with spin-orbit coupling in their ground state. High strain needed to quench phonon processes can be achieved simply by deposition of a thin film \cite{thin.film}, which passively stresses the underlying crystal. A NEMS platform can provide the added benefit of active wavelength tuning, which can enable generation of indistinguishable photons from multiple emitters, and hence scalable photonic quantum networks \cite{SiVcqed, BurekFiber}. Another natural extension of our work is coherent coupling of the SiV spin to a well-defined mechanical mode, which will enable the use of phonons as quantum resource. In particular, we can combine the large strain susceptibility of the SiV \cite{SiVPRB} with mechanical resonators of dimensions close to the phonon wavelength, such as optomechanical crystals \cite{BurekOMC} to obtain orders of magnitude larger spin-phonon interaction strengths compared with previous works \cite{GregHBAR, AniaCantilever, PatrickCantilever, LoncarCantilever, Shimon, Arcizet}, leading to strong spin-phonon coupling. In this regime, one can realise phonon-mediated two-qubit gates \cite{SAWquantum} analogous to those implemented with trapped ions \cite{cirac.zoller}, and achieve quantum non-linearities required to deterministically generate single phonons and non-classical mechanical states \cite{TahanPhonodynamics, Lodahl, PhononCounting,Cleland,Chu.Transmon}, a long sought-after goal since phonons can be used to interface spins with other quantum systems such as superconducting qubits \cite{hybrid}.

\begin{figure}
\includegraphics[width=0.6\columnwidth]{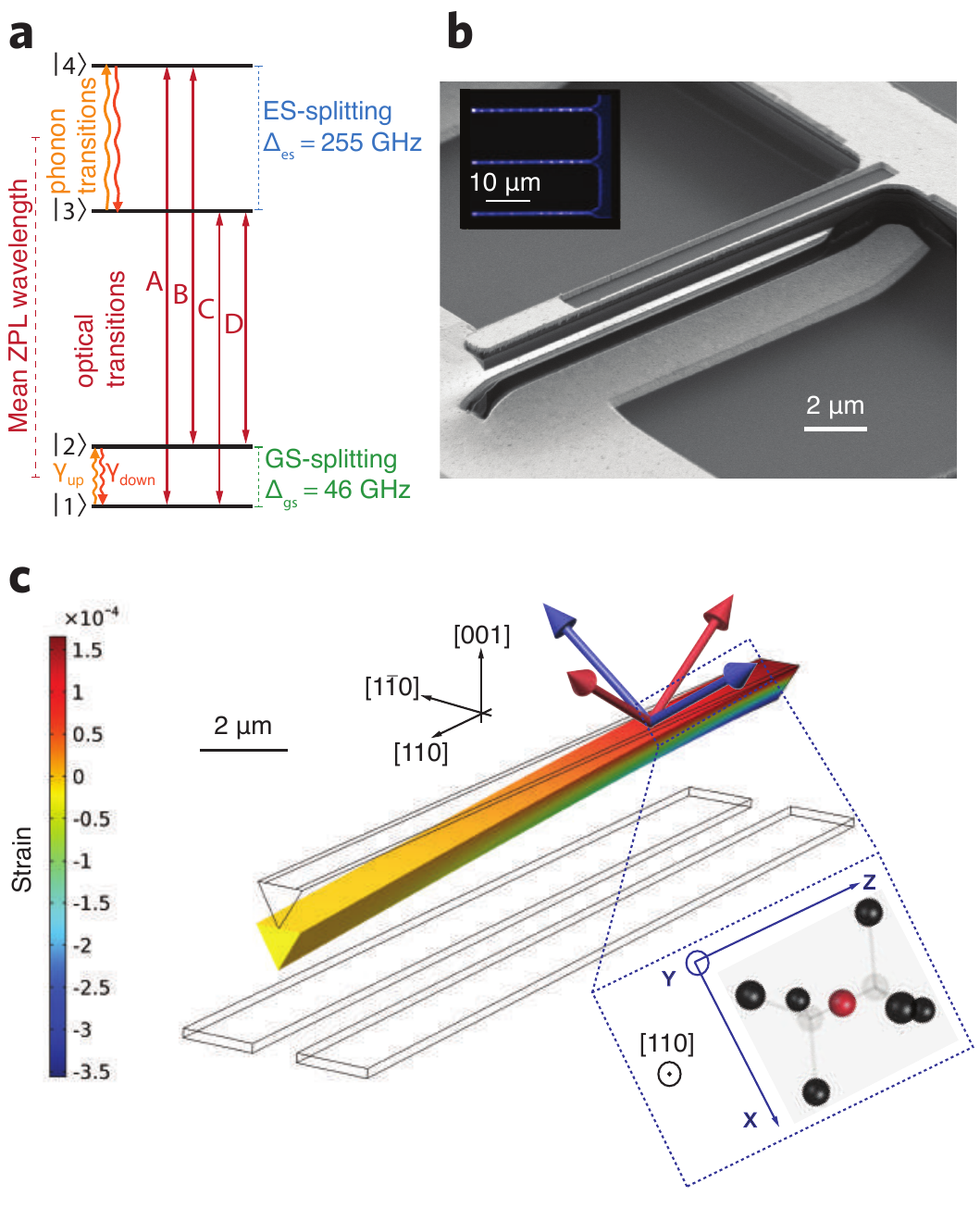}
\caption{(a) Electronic level structure of the SiV showing the mean zero phonon line (ZPL) wavelength, frequency splittings between orbital branches in the ground state (GS) and excited state (ES) ($\Delta_{\mathrm{gs}}$ and $\Delta_{\mathrm{es}}$ respectively) at zero strain, and the four optical transitions A, B, C, and D. Also shown are single-phonon transitions in the GS and ES manifolds. (b) Scanning electron microscope (SEM) image of a representative diamond NEMS cantilever. Dark regions correspond to diamond, and light regions correspond to metal electrodes. (Inset) Confocal photoluminescence image of three adjacent cantilevers. The array of bright spots in each cantilever is fluorescence from SiV centres. (c) Simulation of the displacement of the cantilever due to the application of a DC voltage of 200 V between the top and bottom electrodes. The component of the strain tensor along the long axis of the cantilever is displayed using the colour scale. Crystal axes of diamond are indicated in relation to the geometry of the cantilever. Arrows on top of the cantilever indicate the highest symmetry axes of four possible SiV orientations, and their colour indicates separation into two distinct classes upon application of strain. SiVs studied in this work are shown by blue arrows are oriented along $[1\bar{1}1]$, $[\bar{1}11]$ directions, are orthogonal to the cantilever long-axis, and experience strain predominantly in the plane normal to their highest symmetry axis. Inset shows the molecular structure of such a transverse orientation SiV along with its internal axes, when viewed in the plane normal to the $[110]$ axis.}
\label{fig1}
\end{figure}

\begin{figure}
\includegraphics[width=\columnwidth]{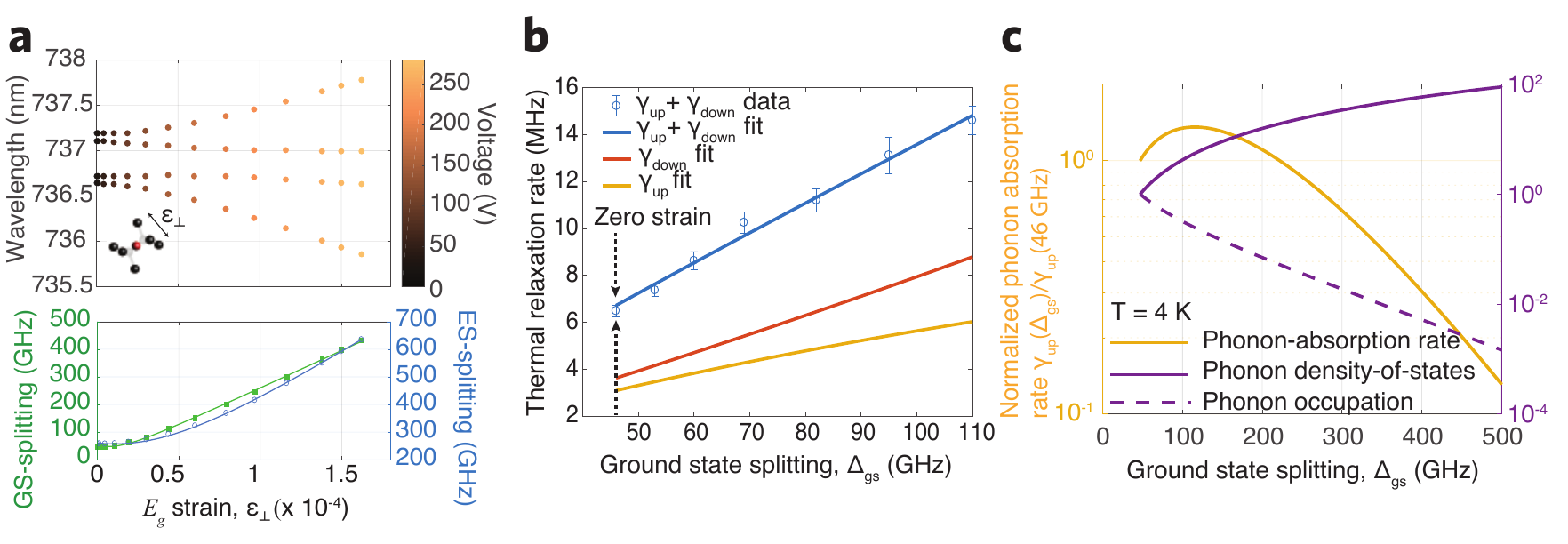}
\caption{(a) Strain response of a transverse orientation SiV labelled with a blue arrow in Fig. 1c. Wavelengths of the four optical transitions A, B, C, and D are recorded against strain. See \cite{SuppInfo} for raw PLE data with applied voltages. The lower panel shows orbital splittings within GS (solid green squares) and ES (open blue circles) extracted from the optical transition wavelengths. Solid curves are fits to group theory based strain response model \cite{SiVPRB, Hepp.thesis}. (b) Thermal relaxation rates between GS orbital branches vs. their energy splitting. Fit to model in \cite{SuppInfo} allows extraction of the phonon-absorption rate $\gamma_{\mathrm{up}}$ and phonon-emission rate $\gamma_{\mathrm{down}}$. (c) Calculated phonon-absorption rate $\gamma_{\mathrm{up}}(\Delta_{\mathrm{gs}})$ (solid yellow line) as a function of GS-orbital splitting $\Delta_{\mathrm{gs}}$ at temperature $T =$ 4K. Left $y$-axis indicates the magnitude of this rate normalized to the value at zero strain, $\gamma_{\mathrm{up}}($46 GHz$)$. Right $y$-axis indicates the two competing factors whose product determines $\gamma_{\mathrm{up}}$: the phonon density of states (normalized to its value at zero strain), shown with the solid violet line, and the thermal occupation of acoustic modes shown with the dashed violet line.}
\label{fig2}
\end{figure}

\begin{figure}
\includegraphics[width=\columnwidth]{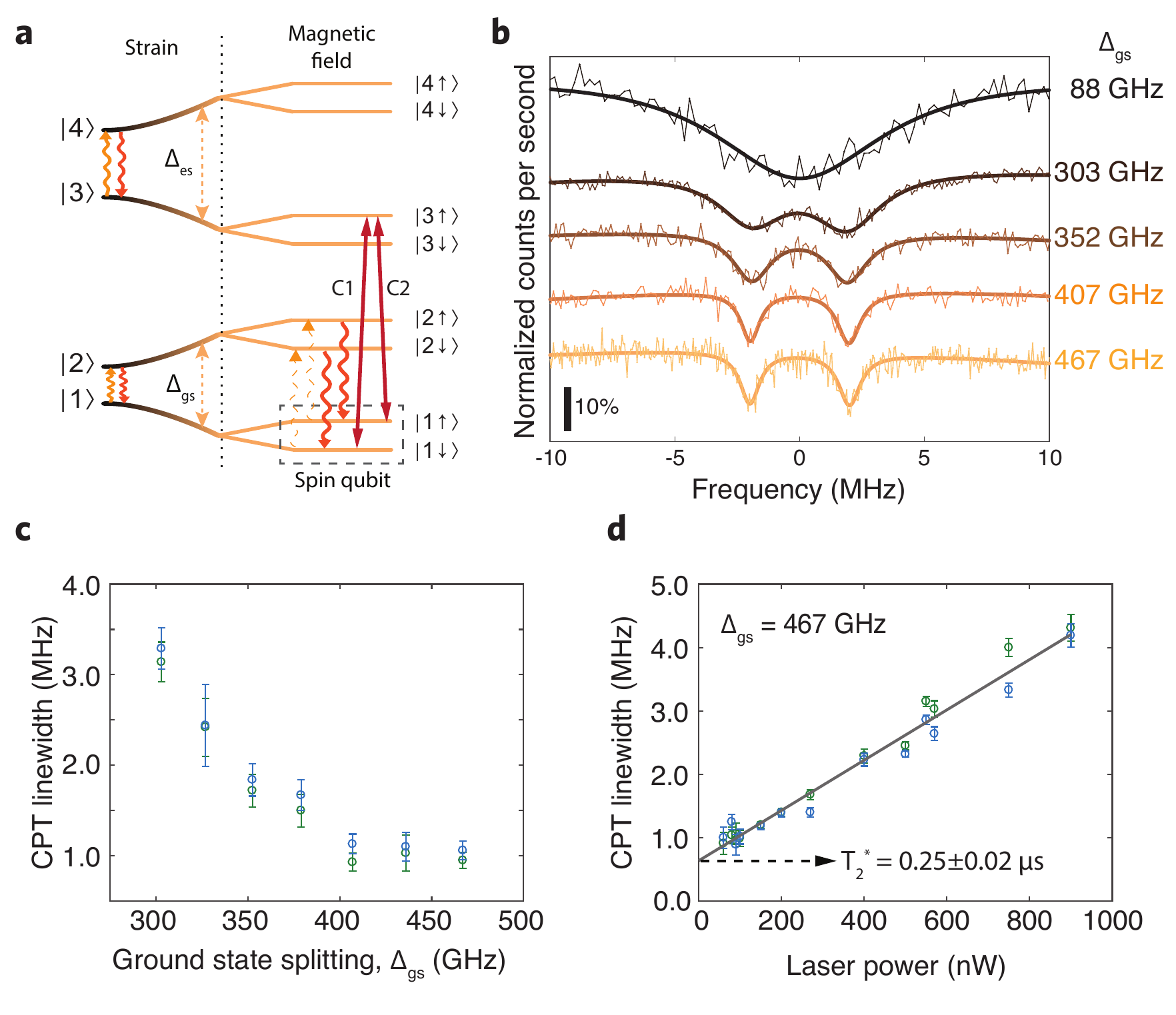}
\caption{(a) SiV level structure in the presence of strain and external magnetic field. A spin qubit is defined with levels $\ket{1\downarrow}$ and $\ket{1\uparrow}$ on the lower orbital branch of the GS. This qubit can be polarized, and prepared optically using the $\Lambda$-scheme provided by transitions C1 and C2. Phonon transitions within ground- and excited-state manifolds are also indicated. The upward phonon transition (phonon absorption process) can be suppressed at high strain, thereby mitigating the effect of phonons on the coherence of the spin qubit. (b) Coherent population trapping (CPT) spectra probing the spin transition at increasing values of the GS orbital splitting $\Delta_{\mathrm{gs}}$ from top to bottom. Bold solid curves are Lorentzian fits. Optical power is adjusted in each measurement to minimize power-broadening. (c) Linewidth of CPT dips (estimated from Lorentzian fits) as a function of GS orbital splitting $\Delta_{\mathrm{gs}}$ indicating improvement in spin coherence with increasing strain. (d) Power dependence of CPT-linewidth at the highest strain condition ($\Delta_{\mathrm{gs}}$=467 GHz). Data points are estimated linewidths from CPT measurements, and the solid curve is a linear fit, which reveals linewidth of $0.64\pm 0.06$ MHz corresponding to $T_2^* = 0.25\pm 0.02 \mathrm{\mu s}$.}
\label{fig4}
\end{figure}

\section*{Acknowledgements}
This work was supported by STC Center for Integrated Quantum Materials (NSF Grant No. DMR-1231319), ONR MURI on Quantum Optomechanics (Award No. N00014-15-1-2761), NSF EFRI ACQUIRE (Award No. 5710004174), the University of Cambridge, the ERC Consolidator Grant PHOENICS, and the EPSRC Quantum Technology Hub NQIT (EP/M013243/1). B.P. thanks Wolfson College (University of Cambridge) for support through a research fellowship. Device fabrication was performed in part at the Center for Nanoscale Systems (CNS), a member of the National Nanotechnology Infrastructure Network (NNIN), which is supported by the National Science Foundation under NSF award no. ECS-0335765. CNS is part of Harvard University. Focused ion beam implantation was performed under the Laboratory Directed Research and Development Program and the Center for Integrated Nanotechnologies, an Office of Science (SC) user facility at Sandia National Laboratories operated for the DOE (contract DE-NA0003525) by National Technology and Engineering Solutions of Sandia, LLC., a wholly owned subsidiary of Honeywell International, Inc. We thank D. Perry for performing the focused ion beam implantation, and K. De Greve and M. W. Doherty for helpful discussions. 



\end{document}